\documentclass[showpacs,preprint,amsmath,amssymb]{revtex4}
\usepackage{amsfonts}

\usepackage{graphicx}
\usepackage{bm}
\newcommand{\beq}{\begin{equation}}
\newcommand{\eeq}{\end{equation}}
\newcommand{\beqa}{\begin{eqnarray}}
\newcommand{\eeqa}{\end{eqnarray}}
\newcommand{\beqar}{\begin{eqnarray*}}
\newcommand{\eeqar}{\end{eqnarray*}}

\newcommand{\ket}[1]{\mbox{$\left|{#1}\right\rangle$}}

\def\I{{\rm i}}

\newcounter{saveeqn}

\begin{document}
\title{Quantum strategy in moving frames}
\author{Jian-Chuan Tan}\email{tanjc@mail.ustc.edu.cn}
\affiliation{Quantum Theory Group, Department of Modern Physics\\
University of Science and Technology of China, Hefei, 230026,
P.R.China}
\author{An Min Wang}\email{anmwang@ustc.edu.cn}\thanks{Corresponding author}
\affiliation{Quantum Theory Group, Department of Modern Physics\\
University of Science and Technology of China, Hefei, 230026,
P.R.China}

\begin{abstract}
We investigate quantum strategy in moving frames by considering
Prisoner's Dilemma and propose four thresholds of $\gamma$ for two
players to determine their \textit{Nash Equilibria}. Specially, an
interesting phenomenon appears in relativistic situation that the
quantum feature of the game would be enhanced and diminished for
different players whose particle's initial spin direction are
respectively parallel and antiparallel to his/her movement
direction, that is, for the former the quantum feature of the game
is enhanced while for the latter the quantum feature would be
diminished. Thus a classical latter could still maintain his/her
strictly dominant strategy (classical strategy) even if the game
itself is highly entangled.
\end{abstract}


\maketitle

Strategy theory (or Game theory) is a branch of applied mathematics
devised to analyze certain situations in which there is an interplay
between parties that may have similar, opposed, or mixed interests.
It draws broad attention because of its practical application in
Economics, Politics, and other fields which involve cooperation or
conflict \cite{Game theory}. As an applied mathematical theory,
strategy theory inevitably possesses its own physical properties. It
is not surprising, since a game should be played through some
strategies, and these strategies must be put in practice to some
physical carriers. Thus the traits of the carriers under some
certain physical conditions would affect the result of a game. Based
on this consideration, to explore how to gain as much as reward in a
game in some particular physical situations has been a popular
research aspect in recent years.

In 1999, Eisert \textit{et al.} proposed a novel model of quantum
game in terms of the famous nonzero sum game--- Prisoners' Dilemma,
in which the physical carriers are two spin-$\frac{1}{2}$ particles,
and players could adopt some unitary quantum operations as
strategies. Although this model was criticized for not possessing
the dominance over a classical game \cite{Enk'}, we find it is
actually go beyond a classical game and worth studying based on the
considerations that it is important for us to distinguish the
difference between the equivalence of payoffs and the equivalence of
strategies, and that to understand the essences of a cooperative
game and a noncooperative game is of high significance in studying a
game with a physical background. More interestingly, a physical
carrier possesses not only quantum traits but also relativistic
ones. So we are concerning on this effect by using Eisert \textit{et
al.}'s model. In this model, two particles (start in a produce state
$\left| C C \right\rangle$) are initially entangled by a gate
$\hat{J}$ to form a pairs of physical carriers of this game, and
then be distributed to two players, Alice and Bob, who independently
chooses a quantum strategy
\begin{equation}
\hat{U}(\theta, \phi) = \left(
\begin{array}{ccc}
e^{\I \phi} \cos{\theta/2} & \sin{\theta/2} \\
-\sin{\theta/2} & e^{-\I \phi} \cos{\theta/2} \\
\end{array} \right),
\end{equation}
with $0 \leq \theta \leq \pi$ and $0 \leq \phi \leq \pi/2$. Finally,
a disentangling gate $\hat{J}^{\dagger}$ is carried out and the
carrier pair is measured in the computational basis. In terms of
game theory, it exists a new \textit{Nash Equilibrium} (NE), that
is, both of the players choose strategy $\hat{Q} = \hat{U} (0,
\pi/2)$, because strategy $\hat{Q}$ has the property of being
\textit{Pareto optimal}, and help players escape the dilemma in
classical game \cite{Game theory, Eisert}.

Let us restrict the physical carriers to be two spin-$\frac{1}{2}$
particles and denote the states of the particles as: $\left|
\frac{1}{2} \right\rangle = \left| \textit{C} \right\rangle =
\left(\begin{array}{c} 1 \\ 0 \end{array}\right)$, and $\left|
-\frac{1}{2} \right\rangle = \left| \textit{D} \right\rangle =
\left(\begin{array}{c} 0 \\ 1 \end{array}\right)$. Meanwhile, an
arbiter is needed to determine each player's payoff by measuring the
state of the two particles with a physical measurement device, and
the principle of the determination is well known to both players.
The players could only gain expected payoff since quantum mechanics
itself is a probabilistic theory. Alice's and Bob's expected payoffs
are given by
\begin{equation}
\begin{array}{c}
\$_A = \textit{r} \textit{P}_{CC} + \textit{p} \textit{P}_{DD}
 + \textit{t} \textit{P}_{DC} + \textit{s} \textit{P}_{CD}, \\
\$_B = \textit{r} \textit{P}_{CC} + \textit{p} \textit{P}_{DD}
+ \textit{s} \textit{P}_{DC} + \textit{t} \textit{P}_{CD}, \\
\end{array}
\end{equation}
where $P_{a b} = \left| \left\langle a b | \psi _f \right\rangle
\right|^2$ ($a,b=C,D$) is the joint probability that the arbiter's
measure device would display $a,b$. We take $\textit{t}=5,
\textit{r}=3, \textit{p}=1$ and $\textit{s}=0$ in this model
\cite{Eisert}. In this game, we assume that the arbiter moves in the
\emph{x} direction, Alice's particle moves in the \emph{z}
direction, and Bob's the -\emph{z} direction. Thus their movements
cause boosts in the direction of \emph{x}, \emph{z}, and -\emph{z},
respectively. Thus, Alice's and Bob's movement directions are
respectively parallel and antiparallel to their particles' initial
spin directions. We denote the boosts with each's rapidity as
$\alpha$ for the arbiter, $\delta _A$ for Alice, and $\delta _B$ for
Bob.

Of course, the arbiter's boost $\alpha$ respect to a player could
also be equivalent to the player emitting the particle to the
arbiter with a rapidity $-\alpha$ (that is, with $\alpha$ in the
-\emph{x} direction). In this case, we could further think that the
arbiter is at rest, and the two players are far away from the
arbiter, so they have to take part in this game by emitting their
own particles to the arbiter, and the rapidity of each particle will
sort of determine how much payoff the players would attain. Thus, at
what speed the particle is emitted could be controlled by the
player, and we name this speed-control as a relativistic operation.
From our point of view, this model should be worth studying since it
is a well guidance to long-distance games, and even in the near
future when interstellar travel comes true, this model would also be
useful.

Now we set out our game model and its process is illustrated in
Fig.1, in which the Lorentz boost is introduced in
Refs.\cite{Alsing,Ahn}, and $\gamma$ is a monotonic function with
the measure of entanglement, indicating how much the two particles
entangle. The degree of entanglement between the two particles would
decrease if their momentum have distributions, say, with width. So
tracing out the momentum from the Lorentz-transformation density
matrix destroys some of the entanglement \cite{Gingrich}. We assume
the momentum of both particles to be exact, namely no distributions,
thus their degree of entanglement would remain invariant under
Lorentz transformation, and so does $\gamma$. When $\gamma = 0$, the
game's players are separable and the game does not display any
features which go beyond the classical game.

\begin{figure}[htbp]
\centerline{\includegraphics[scale=0.50]{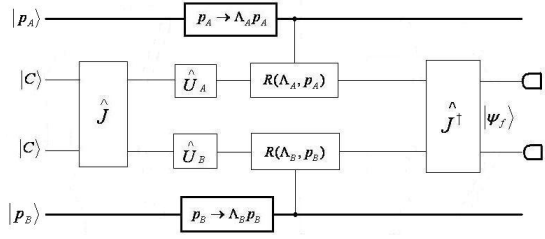}}
\caption{Process of the game model. $\hat{J} = \exp(\I\gamma \hat{D}
\otimes \hat{D}/2)$, $\gamma \in \left[0,\pi/2 \right]$,
$\hat{D}=\hat{U}(\pi, 0)$, is defined in \cite{Eisert} to make the
two particles entangle. $R(\Lambda,\bm{p})$ is the Wigner rotation
applied to a particle. $\hat{U}_A$ and $\hat{U}_B$ are operations
Alice and Bob applies to her and his own particle respectively. }
\label{fig1} \vskip -0.10in
\end{figure}

The Lorentz transformation $\Lambda$ results in a unitary
transformation on states in the Hilbert space that $ \ket{\Psi}
\rightarrow U(\Lambda) \ket{\Psi} $. Thus, the state of entangled
particles under the Lorentz transformation is given by
\beq U(\Lambda) (\hat{U}_A \otimes \hat{U}_B) \hat{J} \left|
\bm{p}_A,\!C;\bm{p}_B,\! C \right\rangle =
\sum_{a,b=C,D}k_{ab}\ket{\psi_{ab}},\eeq
\begin{eqnarray}
\ket{\psi_{CC}}\!\!&=&\!\!\sqrt{\frac{(\Lambda
\bm{p}_A)^0}{\bm{p}_A^0}} \sqrt{\frac{(\Lambda
\bm{p}_B)^0}{\bm{p}_B^0}}\sum_{\sigma,\sigma'}
{D}_{\sigma,\frac{1}{2}}^{(\frac{1}{2})}\left( \emph{R}(\Lambda_A)
\right)\nonumber\\ & &
{D}_{\sigma',\frac{1}{2}}^{(\frac{1}{2})}\left( \emph{R}(\Lambda_B)
\right)a^\dagger (\bm{p}_{A_\Lambda},\sigma)
a^\dagger(\bm{p}_{B_\Lambda},\sigma')\ket{\psi_0},\\
\ket{\psi_{CD}}\!\!&=&\!\!\sqrt{\frac{(\Lambda
\bm{p}_A)^0}{\bm{p}_A^0}} \sqrt{\frac{(\Lambda
\bm{p}_B)^0}{\bm{p}_B^0}}\sum_{\sigma,\sigma'}
{D}_{\sigma,\frac{1}{2}}^{(\frac{1}{2})}\left( \emph{R}(\Lambda_A)
\right)\nonumber\\ & &
{D}_{\sigma',-\frac{1}{2}}^{(\frac{1}{2})}\left( \emph{R}(\Lambda_B)
\right) a^\dagger (\bm{p}_{A_\Lambda},\sigma) a^\dagger
(\bm{p}_{B_\Lambda},\sigma')\ket{\psi_0},\eeqa\beqa
\ket{\psi_{DC}}\!\!&=&\!\!\sqrt{\frac{(\Lambda
\bm{p}_A)^0}{\bm{p}_A^0}} \sqrt{\frac{(\Lambda
\bm{p}_B)^0}{\bm{p}_B^0}}\sum_{\sigma,\sigma'}
{D}_{\sigma,-\frac{1}{2}}^{(\frac{1}{2})}\left( \emph{R}(\Lambda_A)
\right)\nonumber\\ & &
{D}_{\sigma',\frac{1}{2}}^{(\frac{1}{2})}\left( \emph{R}(\Lambda_B)
\right) a^\dagger (\bm{p}_{A_\Lambda},\sigma) a^\dagger
(\bm{p}_{B_\Lambda},\sigma')\ket{\psi_0},\\
\ket{\psi_{DD}}\!\!&=&\!\!\sqrt{\frac{(\Lambda
\bm{p}_A)^0}{\bm{p}_A^0}} \sqrt{\frac{(\Lambda
\bm{p}_B)^0}{\bm{p}_B^0}} \sum_{\sigma,\sigma'}
{D}_{\sigma,-\frac{1}{2}}^{(\frac{1}{2})}\left( \emph{R}(\Lambda_A)
\right)\nonumber\\ & &
{D}_{\sigma',-\frac{1}{2}}^{(\frac{1}{2})}\left( \emph{R}(\Lambda_B)
\right) a^\dagger (\bm{p}_{A_\Lambda},\sigma) a^\dagger
(\bm{p}_{B_\Lambda},\sigma')\ket{\psi_0}.
\end{eqnarray}
where $\ket{\psi_0}$ is the Lorentz invariant vacuum state, and
\beqa k_{CC} &=& e^{\I(\phi_A + \phi_B)} c_{\theta_A} c_{\theta_B}
c_\gamma + \I s_{\theta_A} s_{\theta_B} s_\gamma,\\
k_{CD} &=& -e^{\I\phi_A} c_{\theta_A} s_{\theta_B} c_\gamma + \I
e^{-\I\phi_B} s_{\theta_A} c_{\theta_B} s_\gamma, \\
k_{DC} &=& -e^{\I\phi_B} s_{\theta_A} c_{\theta_B} c_\gamma
 + \I e^{-\I\phi_A} c_{\theta_A} s_{\theta_B} s_\gamma,\\
k_{DD} &=& s_{\theta_A} s_{\theta_B} c_\gamma + \I e^{-\I(\phi_A +
\phi_B)} c_{\theta_A} c_{\theta_B} s_\gamma. \eeqa For simplicity,
we denote \beq c_{x} = \cos \frac{x}{2},\quad s_{x} = \sin
\frac{x}{2}, \eeq where $x$ can be taken as $\theta_A$, $\theta_B$
and $\gamma$ as well as so-called Wigner angle $\Omega_A$ and
$\Omega_B$ respectively with Alice's and Bob's particles. Note that
a particle's Wigner angle is determined by the rapidities of itself
($\delta$) and the arbiter ($\alpha$) \cite{Alsing}\cite{Ahn},
\begin{equation}
\Omega_\tau = \arctan \frac{\sinh \alpha \sinh \delta_\tau}{\cosh
\alpha + \cosh \delta_\tau}, \tau = A, B.
\end{equation}

The final state measured by the arbiter is $\left| \psi_f
\right\rangle = \hat{J}^{\dagger} U(\Lambda) (\hat{U}_A \otimes
\hat{U}_B) \hat{J} \left| \bm{p}_A, C; \bm{p}_B, C \right\rangle$.
We have
\begin{equation}
\left( \begin{array}{c} \mathfrak{p}_1\\ \mathfrak{p}_2\\ \mathfrak{p}_3\\
\mathfrak{p}_4
\end {array} \right) = \left(
\begin{array}{cccc}
\omega_1 & \omega^*_2 & -\omega^*_3 & -\omega_4 \\
-\omega^*_2 & \omega_1 & \omega_4 & -\omega_3 \\
\omega^*_3 & \omega_4 & \omega_1 & -\omega_2 \\
-\omega_4 & \omega^*_3 & -\omega^*_2 & \omega_1
\end{array} \right) \left(
\begin{array}{c} k_{CC}\\ k_{CD}\\ k_{DC}\\ k_{DD}
\end{array} \right),
\end{equation}
where $\omega_1 = c_\gamma c_{\Omega_A} c_{\Omega_B} + \I s_\gamma
s_{\Omega_A} s_{\Omega_B}$, $\omega_2 = c_\gamma c_{\Omega_A}
s_{\Omega_B} + \I s_\gamma s_{\Omega_A} c_{\Omega_B}$, $\omega_3 =
c_\gamma s_{\Omega_A} c_{\Omega_B} + \I s_\gamma c_{\Omega_A}
s_{\Omega_B}$, and $\omega_4 = c_\gamma s_{\Omega_A} s_{\Omega_B} +
\I s_\gamma c_{\Omega_A} c_{\Omega_B}$, and $*$ denotes complex
conjugation. Thus we get ${P}_{CC}= \left|\mathfrak{p}_1\right|^2$,
${P}_{CD}= \left|\mathfrak{p}_2\right|^2$, ${P}_{DC}=
\left|\mathfrak{p}_3\right|^2$, and ${P}_{DD}=
\left|\mathfrak{p}_4\right|^2$.

Actually, how much the two particles are initially entangled would
be essential to this game model, since $\gamma$ induces some
features which go beyond the classical game. Du \textit{et al.}
found two thresholds of $\gamma$ in the Quantum Prisoners'
Dilemma--- $\gamma _{th1} = \arcsin {\sqrt{1/5}}$ and $\gamma _{th2}
= \arcsin{\sqrt{2/5}}$, which separate the game into three regions:
classical region ($\gamma \in \left[0, \gamma _{th1} \right)$),
intermediate region ($\gamma \in \left[\gamma _{th1}, \gamma _{th2}
\right)$), and fully quantum region ($\gamma \in \left[\gamma
_{th2}, \pi/2 \right]$), see Ref.\cite{Du} \cite{Du'}. According to
Du, the classical region means in this domain, the game behaves
classically, i.e., the NE of the game is $\hat{D}\otimes\hat{D}$; in
the quantum region, the game is similar to the maximally entangled
one in Eisert's Letter \cite{Eisert} that $\hat{Q}\otimes\hat{Q}$
becomes the new NE and has the property to be \emph{Pareto Optimal};
while the intermediate region possesses compatibility to $\hat{D}$
and $\hat{Q}$, where $\hat{D}\otimes\hat{D}$ is no longer the NE
because each player could improve his/her payoff by unilaterally
deviating from the strategy $\hat{D}$, thus two \textit{Nash
Equilibria} (NE's) $\hat{D}\otimes\hat{Q}$ and
$\hat{Q}\otimes\hat{D}$ emerge \cite{Du}.

In order to explore the relativistic-quantum features of this game,
we take four situations as examples, in which $4$-kinds of payoffs
are considered for each player--- (a) Alice moves at low speed (AL)
\& Bob moves at low speed (BL), (b) Alice moves at low speed (AL) \&
Bob moves at high speed (BH), (c) Alice moves at high speed (AH) \&
Bob moves at low speed (BL), and (d) Alice moves at high speed (AH)
\& Bob moves at high speed (BH); and $G_1 :=
\$(\hat{D}\otimes\hat{D})$, $G_2 := \$(\hat{Q}\otimes\hat{D})$, $G_3
:= \$(\hat{D}\otimes\hat{Q})$, and $G_4 :=
\$(\hat{Q}\otimes\hat{Q})$. And we concentrate our discussion to a
simple but typical strategy set $S = \{\hat{D},\hat{Q}\}$, since
$\hat{D} = \hat{U}(\pi, 0)$ is a classical spin-rotating operation
which could be implemented by sort of classical equipments, while
$\hat{Q} = \hat{U}(0, \pi/2)$ is a purely phase-controlling
operation which could only be implemented by a quantum gate. It is
an essential difference between these two strategies. Thus, there
are at most six thresholds of $\gamma$ ($\gamma_{\mu\nu},
\mu,\nu=1,2,3,4$, with $\mu<\nu$, where $\gamma_{\mu\nu}$ is the
point where $G_\mu = G_\nu$) for each player's payoff in each
situation. Among these $\gamma_{\mu\nu}$, there are two thresholds
are essential for each player--- for Alice, they are $\gamma_{12}$
and $\gamma_{34}$, we denote them as $\gamma^A_{12}$ and
$\gamma^A_{34}$; similarly, for Bob, they are $\gamma^B_{13}$ and
$\gamma^B_{24}$. These four thresholds are essential because they
demonstrate Alice's and Bob's strictly dominant strategies (SDS) for
different $\gamma \in \left[ 0, \frac{\pi}{2} \right]$ \cite{Game
theory}. Fig.2 illustrates Alice's and Bob's payoffs in the four
situations. Here, as for low and high speed we can respectively take
$\Omega_\tau = \frac{\pi}{16}$ and $\Omega_\tau = \frac{7\pi}{16}$.
As is mentioned in Ref.\cite{Ahn}, $\Omega_\tau$ is a monotonic
function with player $\tau$'s and the arbiter's speeds. Thus in this
example, $\Omega_\tau = \frac{\pi}{16}$ corresponds to arbiter's
speed 0.01c and $\tau$'s speed 0.001c, while $\Omega_\tau =
\frac{7\pi}{16}$ corresponds to arbiter's speed 0.97c and $\tau$'s
speed 0.908c, where c is the light-speed. The arbiter's speed is
equivalent to the same speed that the player emits his/her particle
in the -\emph{x} direction, as mentioned above.

\begin{figure}[h]
\leftline{\includegraphics[scale=0.5]{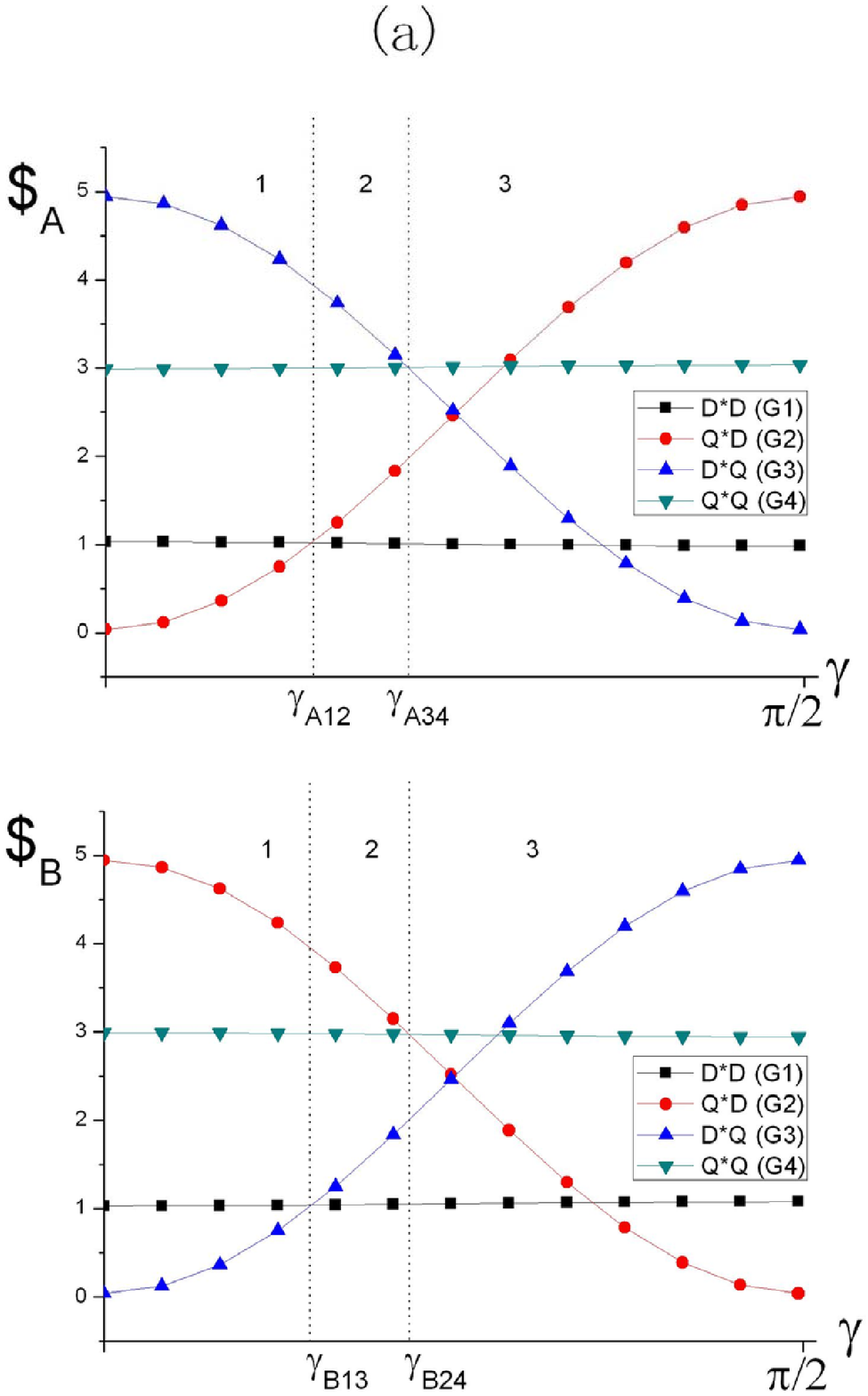}
\includegraphics[scale=0.5]{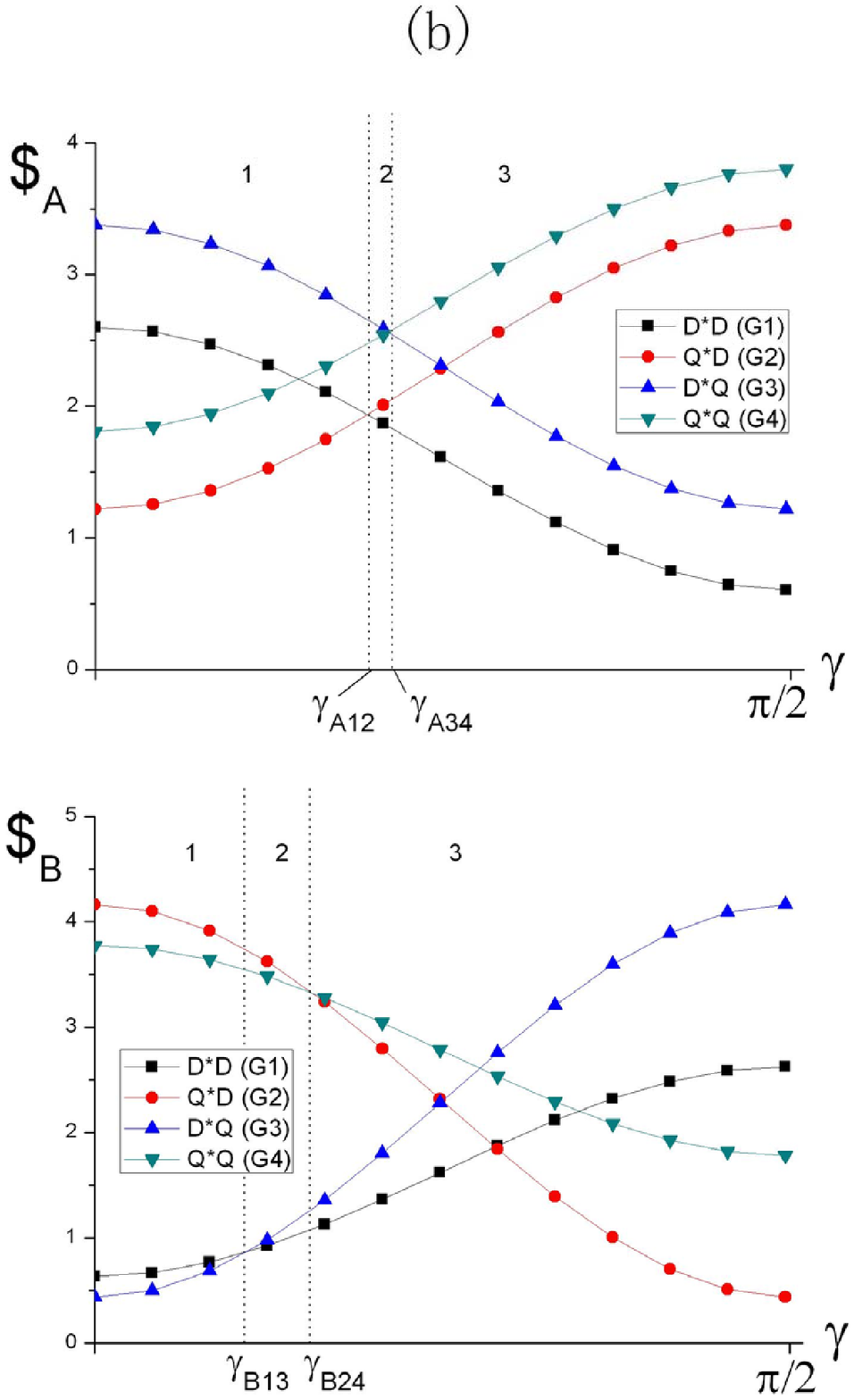}}
\leftline{\includegraphics[scale=0.5]{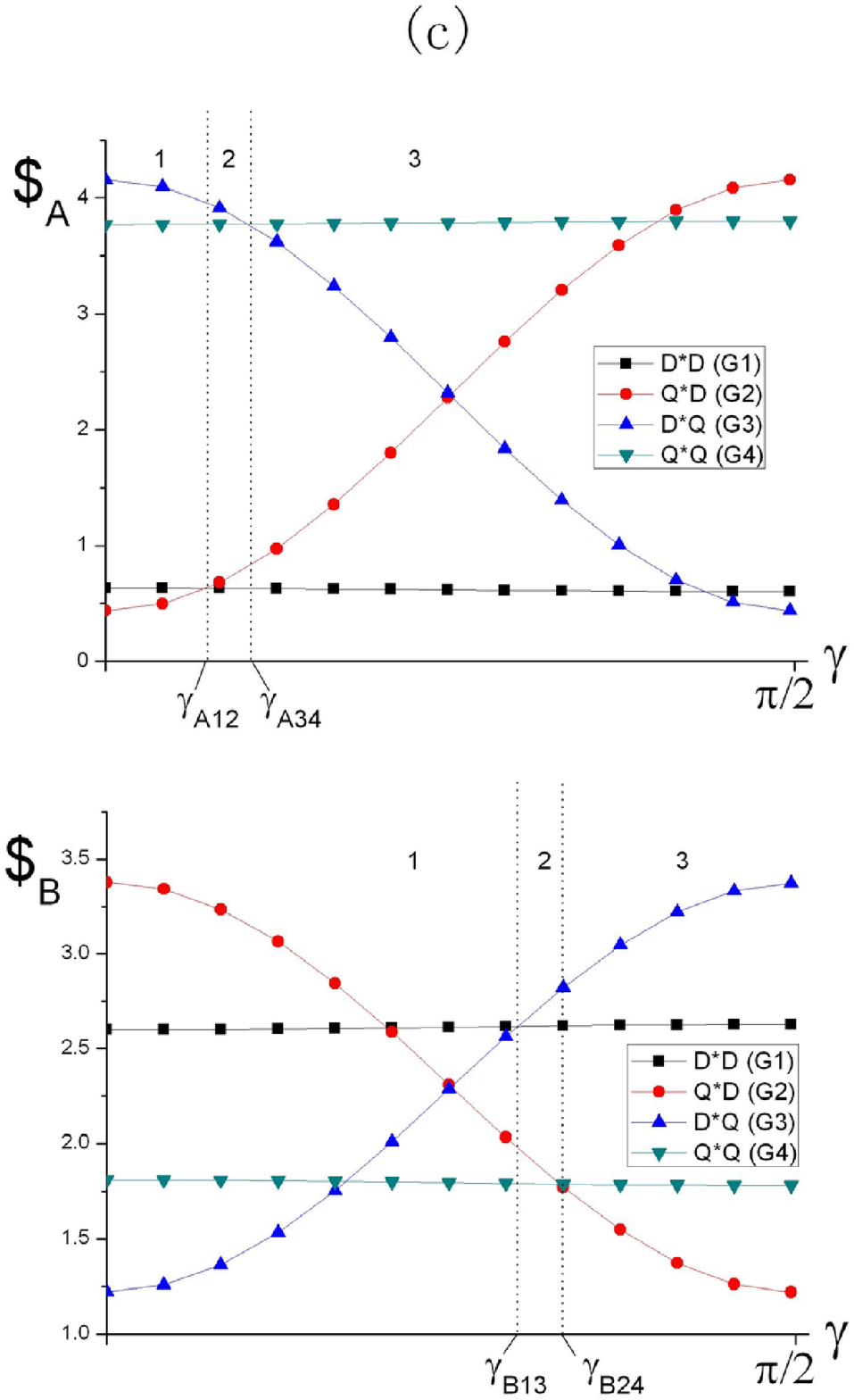}
\includegraphics[scale=0.5]{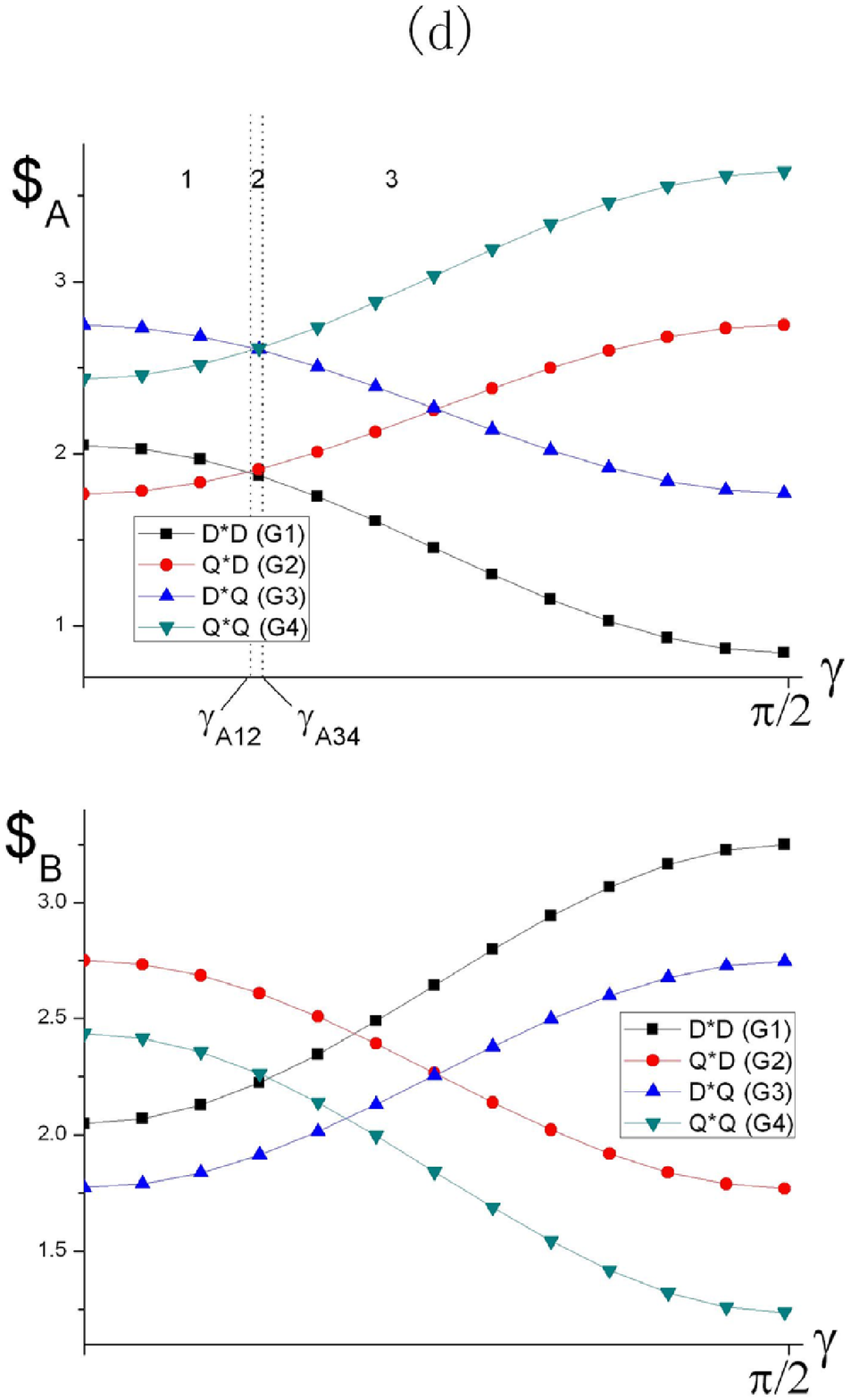}}
\caption{(Color online) Alice's and Bob's payoffs in 4 situations---
(a) AL \& BL, (b) AL \& BH, (c) AH \& BL, and (d) AH \& BH.}
\label{fig2}
\end{figure}

In Fig.2, we name the region where $\gamma^A_{12} < \gamma <
\gamma^A_{34}$ Alice's transition region ($\mathcal{T}_A$), and
where $\gamma^B_{13} < \gamma < \gamma^B_{24}$ Bob's transition
region ($\mathcal{T}_B$). If $\gamma$ is on the left side of
$\mathcal{T}_\tau$, then $\tau$'s SDS is $\hat{D}$ (purely classical
strategy); if $\gamma$ is on the right side of $\mathcal{T}_\tau$,
the SDS is $\hat{Q}$ (purely quantum strategy); while if $\gamma$ is
in $\mathcal{T}_\tau$, $\tau$ would have no SDS, but the NE still
exist. Game theory proves that the combination of each player's SDS
must be the NE of the game, but a NE may not be the combination of
each's SDS \cite{Game theory}. From Fig.2, we could see that in some
situations, $\mathcal{T}_A$ and $\mathcal{T}_B$ overlap partially
with each other, and in the overlapping region, two new NE's
$\hat{D}\otimes\hat{Q}$ and $\hat{Q}\otimes\hat{D}$ appear, although
there is no SDS exists for each player. On the other hand, if
$\gamma$ is in $\mathcal{T}_A$ but not in $\mathcal{T}_B$, Bob has
SDS $\hat{D}$ or $\hat{Q}$, but Alice has not, in this case, the NE
is $\hat{Q}\otimes\hat{D}$ or $\hat{D}\otimes\hat{Q}$, that is to
say, Alice should choose the strategy opposite to Bob's SDS. It is
similar to the case that $\gamma$ is in $\mathcal{T}_B$ but not in
$\mathcal{T}_A$. What is noteworthy is the highly relativistic
situation in Fig.2.(d): $\Omega_A = \Omega_B = \frac{7\pi}{16}$. In
this case, there is no transition region for Bob, and for all
$\gamma \in \left[ 0, \frac{\pi}{2} \right]$, Bob's SDS is
$\hat{D}$, that is to say, when Alice's and Bob's particles both
move at very high speed, the game behaves classically for Bob, even
if he is highly entangled with Alice. It is an interesting
phenomenon that the relativistic operations would diminish the
quantum feature of the game. Fig.3 shows the area where Bob's SDS is
$\hat{D}$ for all $\gamma \in \left[ 0, \frac{\pi}{2} \right]$,
i.e., where the relativistic operation entirely eliminate the
quantum feature of the game for Bob.

\begin{figure}[htbp]
\centerline{\includegraphics[scale=0.50]{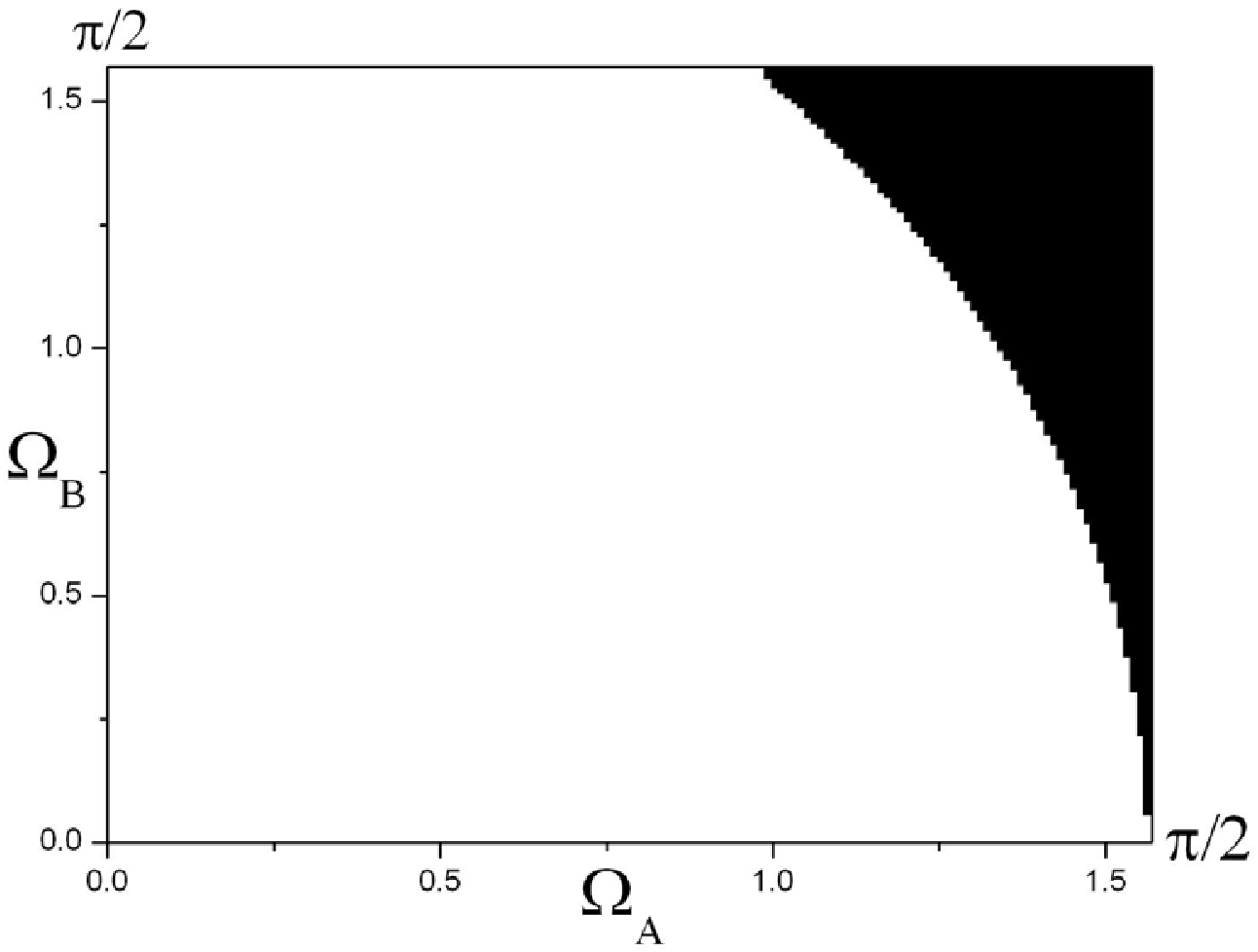}} \caption{The
shadowed area indicates the situation in which Bob's SDS is always
$\hat{D}$ in spite of how much the two particles are entangled.}
\label{fig3}
\end{figure}

In fact, the four thresholds vary with $\Omega_A$ and $\Omega_B$ as
\begin{eqnarray}
\gamma^A_{12}\! = \! \arcsin\!{\sqrt{\frac{c^2_{\Omega_A}
c^2_{\Omega_B}
 \!-\! 2 s^2_{\Omega_A} s^2_{\Omega_B}\! + \!2 c^2_{\Omega_A} s^2_{\Omega_B}
 \!-\! s^2_{\Omega_A} c^2_{\Omega_B}}{5 c^2_{\Omega_A} c^2_{\Omega_B}
 \!-\! 5 s^2_{\Omega_A} s^2_{\Omega_B}\! +\!
 3 c^2_{\Omega_A} s^2_{\Omega_B} \!+ \!2 s^2_{\Omega_A}
 c^2_{\Omega_B}}}},\\
\gamma^A_{34}\! = \! \arcsin\!{\sqrt{\frac{2 c^2_{\Omega_A}
c^2_{\Omega_B}\! -\! s^2_{\Omega_A} s^2_{\Omega_B} \!+\!
c^2_{\Omega_A} s^2_{\Omega_B}\! -\! 2 s^2_{\Omega_A}
c^2_{\Omega_B}}{5 c^2_{\Omega_A} c^2_{\Omega_B} \!-\! 5
s^2_{\Omega_A} s^2_{\Omega_B}\! + \!3 c^2_{\Omega_A}
s^2_{\Omega_B}\! +\! 2 s^2_{\Omega_A} c^2_{\Omega_B}}}}, \\
\gamma^B_{13}\! = \! \arcsin\!{\sqrt{\frac{c^2_{\Omega_A}
c^2_{\Omega_B} \!- \!2 s^2_{\Omega_A} s^2_{\Omega_B} \!-\!
c^2_{\Omega_A} s^2_{\Omega_B}\! +\! 2 s^2_{\Omega_A}
c^2_{\Omega_B}}{5 c^2_{\Omega_A} c^2_{\Omega_B}\! -\! 5
s^2_{\Omega_A} s^2_{\Omega_B} \!-\! 3 c^2_{\Omega_A} s^2_{\Omega_B}
\!-\! 2 s^2_{\Omega_A} c^2_{\Omega_B}}}},\\
\gamma^B_{24}\! = \! \arcsin\!{\sqrt{\frac{2 c^2_{\Omega_A}
c^2_{\Omega_B}
 \!-\! s^2_{\Omega_A} s^2_{\Omega_B}\! -\! 2 c^2_{\Omega_A} s^2_{\Omega_B}
 \!+\! s^2_{\Omega_A} c^2_{\Omega_B}}{5 c^2_{\Omega_A} c^2_{\Omega_B}
 \!-\! 5 s^2_{\Omega_A} s^2_{\Omega_B}\! -\! 3 c^2_{\Omega_A} s^2_{\Omega_B}
\! -\! 2 s^2_{\Omega_A} c^2_{\Omega_B}}}}.
\end{eqnarray}
always with $\gamma^A_{12} < \gamma^A_{34}$ and $\gamma^B_{13} <
\gamma^B_{24}$. We plot these four thresholds in Fig.4. In
particular, when Alice, Bob and the arbiter are all at rest, i.e.,
$\Omega_A = \Omega_B = 0$, $\mathcal{T}_A$ and $\mathcal{T}_B$
overlap entirely with each other. In this case, $\gamma^A_{12} =
\gamma^B_{13} = \gamma_{th1}$ in Du's paper \cite{Du}, and
$\gamma^A_{34} = \gamma^B_{24} = \gamma_{th2}$, thus two NE's emerge
in the overlapping region.
\begin{figure}[htb]
\leftline{\includegraphics[scale=0.5]{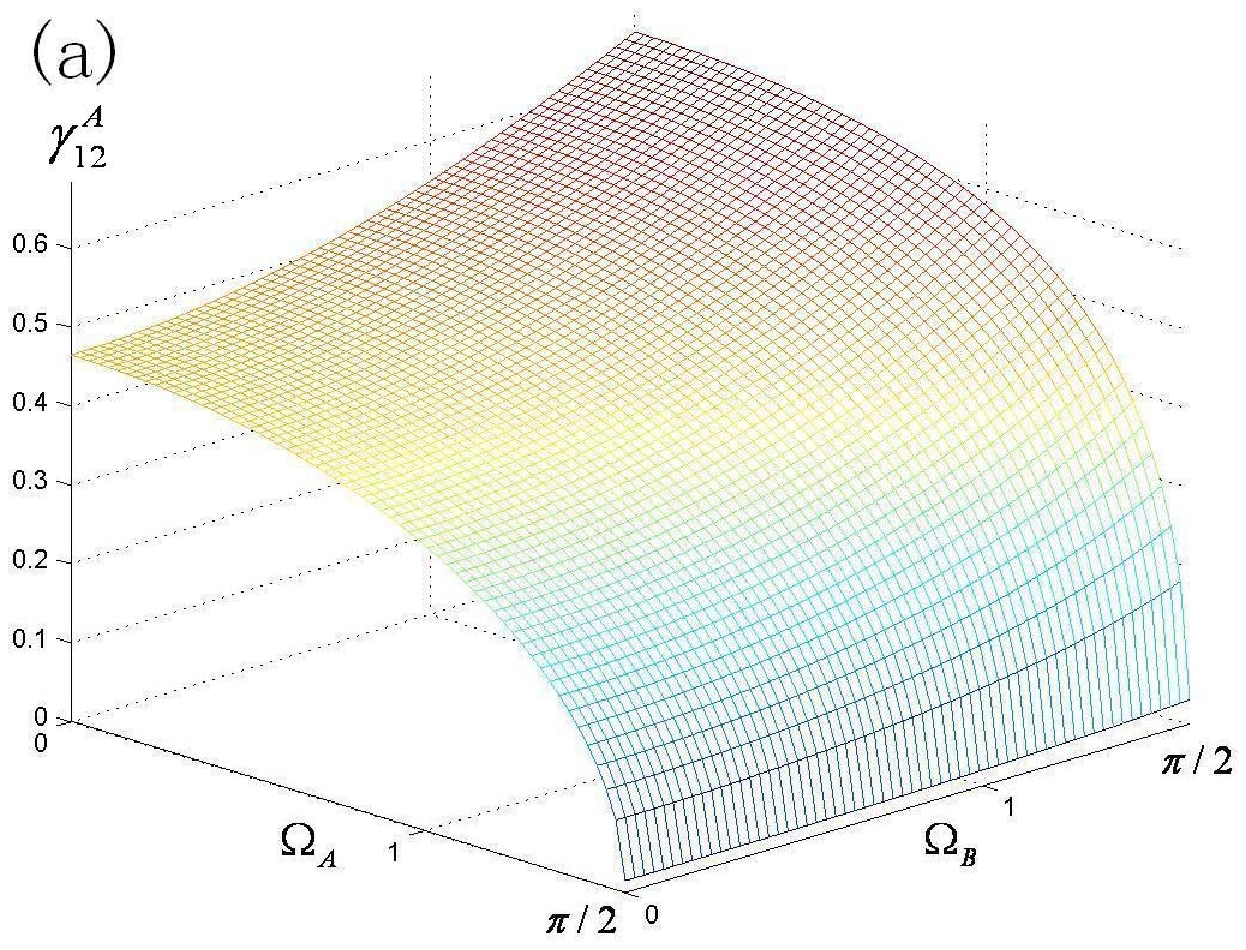}
\includegraphics[scale=0.5]{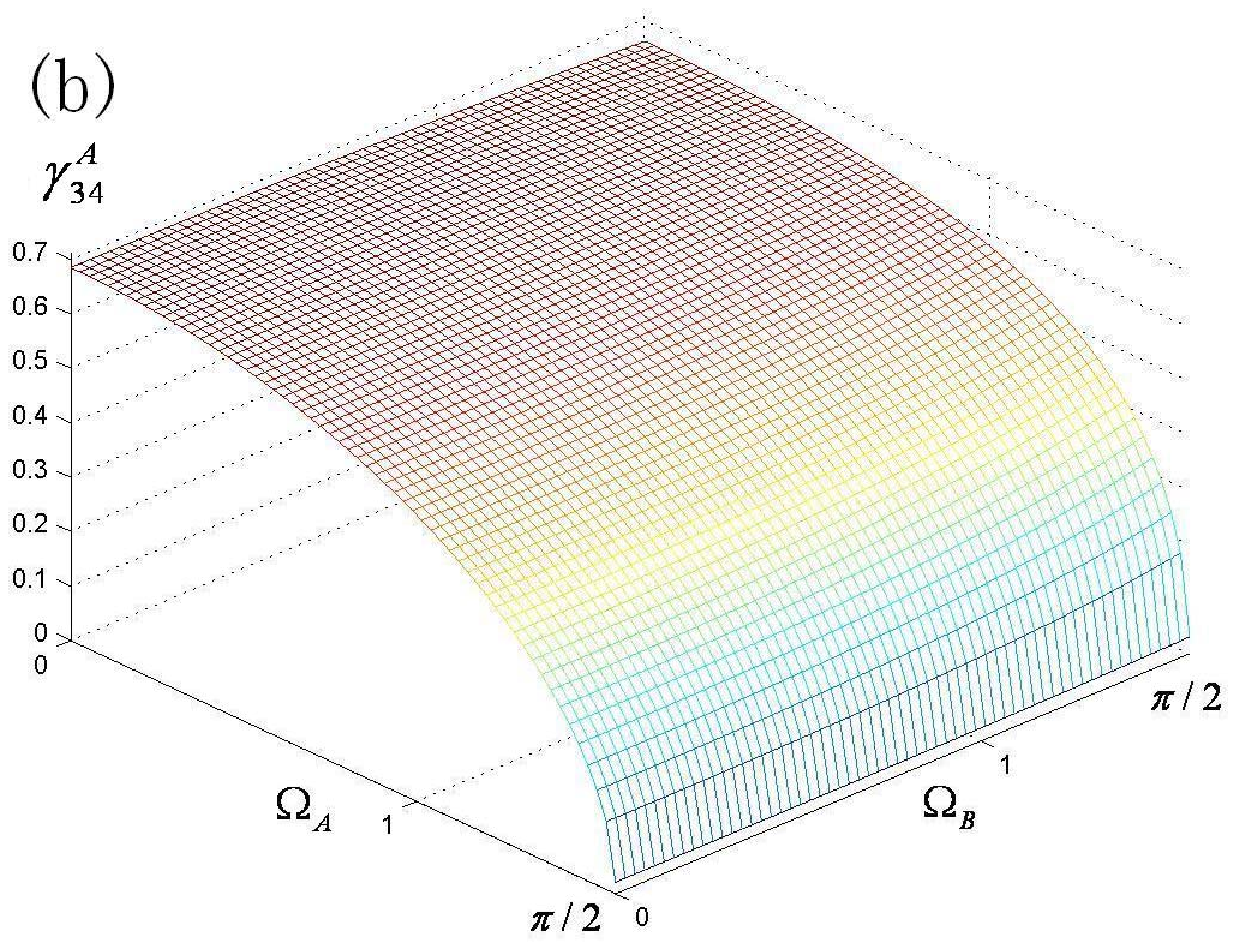}}
\leftline{\includegraphics[scale=0.5]{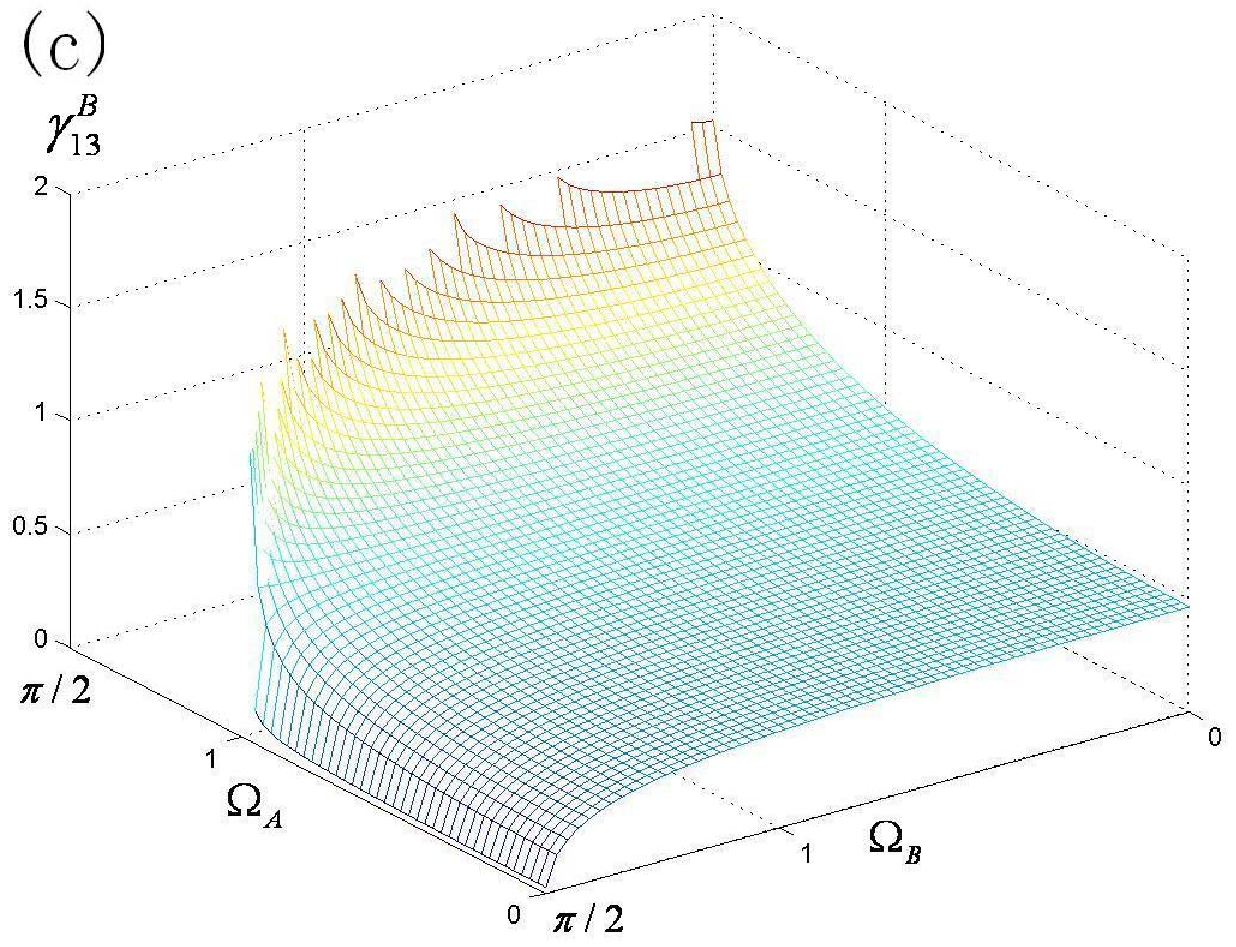}
\includegraphics[scale=0.5]{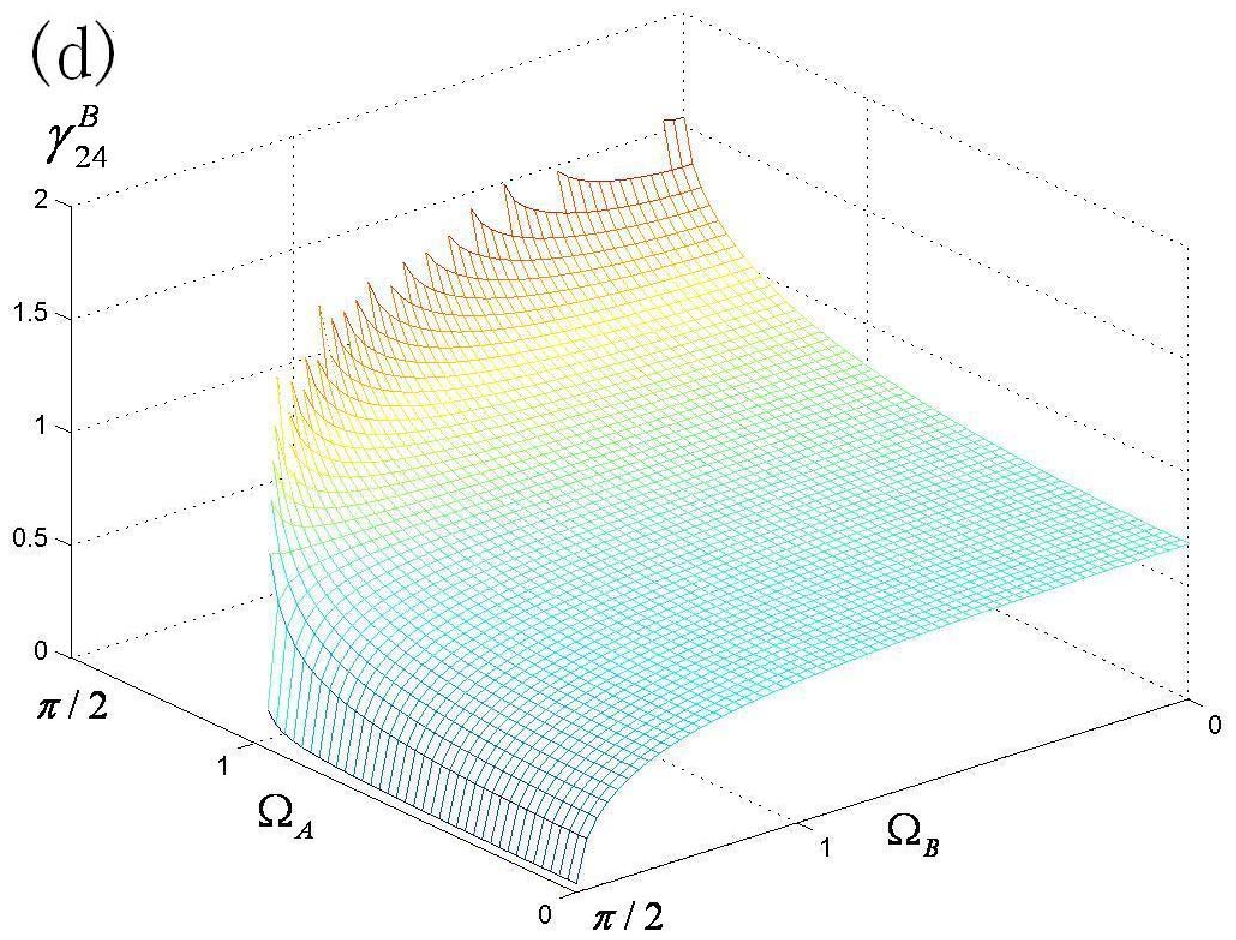}}
\caption{The four thresholds $\gamma^A_{12}$, $\gamma^A_{34}$,
$\gamma^B_{13}$ and $\gamma^B_{24}$, which divide the game into
three regions respectively according to $\gamma$, and determine the
\textit{Nash Equilibrim} of this game.}\label{fig4}
\end{figure}

Finally, we could see in Fig.4.(b) that for Alice, $\gamma^A_{34} <
\frac{\pi}{2}$ in all situations, and $\gamma^A_{34} \rightarrow 0$
when $\Omega_A \rightarrow \frac{\pi}{2}$, i.e., when Alice's
particle moves at very high speed, her SDS would be $\hat{Q}$ even
if the two particles are entirely separable; while in Fig.4.(c),
$\gamma^B_{13} > \frac{\pi}{2}$ in some situations, where the
quantum feature of the game is entirely eliminated for Bob, so his
SDS is $\hat{D}$ even if the two particles are entirely entangled.
That is to say, in the same game, the relativistic operations
enhance the quantum feature of the game for Alice, but diminish it
for Bob.

In summary, we have demonstrated that some new and interesting
features appear if classical games such as Prisoners' Dilemma are
extended to the quantum and relativistic domain, in which the
initial symmetry of this game is broken by the respect movements of
the two players. We also propose four thresholds for Alice and Bob,
which divide the game into three regions in which different strictly
dominant strategies emerge, and how \textit{Nash Equilibrium} is
determined in different situations. Moreover, a interesting
phenomenon appears in relativistic situation that the relativistic
operations could enhance the quantum feature of the game for the
player whose particle's initial spin direction is parallel to its
movement direction (Alice), but diminish it for the one whose
particle's initial spin direction is antiparallel to its movement
direction (Bob), i.e., the respect movements of Alice, Bob and the
arbiter determine ``how quantum" the game is for each player. We
believe these properties would be useful to guide remote games in
the future and that extending game theory to quantum and
relativistic domain would lead us to understand the physical essence
of game theory.

We are grateful to all the collaborators of our quantum theory group
in the institute for theoretical physics of our university. We thank
Prof. Lewenstein and Zeyang Liao for triggering and useful
discussion. This work was supported by the National Natural Science
Foundation of China under Grant No. 60573008.

\end{document}